# Information Retrieval and Classification of Real-Time Multi-Source Hurricane Evacuation Notices

Tingting Zhao[1*], Shubo Tian[2], Jordan Daly[3], Melissa Geiger[3], Minna Jia[4], Jinfeng Zhang[2*]

[1] Department of Geography, Florida State University, Tallahassee, FL, USA
[2] Department of Statistics, Florida State University, Tallahassee, FL, USA
[3] Department of Earth, Ocean, and Atmospheric Science, Florida State University, Tallahassee, FL, USA
[4] FSU Survey Foundry, Florida State University, Tallahassee, FL, USA

* Corresponding authors: Tingting Zhao tzhao@fsu.edu; Jinfeng Zhang jinfeng@stat.fsu.edu

Declarations of interest: none

This work was supported by National Institutes of Health (NIH) grant # R01GM126558

**Abstract**

For an approaching disaster, the tracking of time-sensitive critical information such as hurricane evacuation notices is challenging in the United States. These notices are issued and distributed rapidly by numerous local authorities that may spread across multiple states. They often undergo frequent updates and are distributed through diverse online portals lacking standard formats. In this study, we developed an approach to timely detect and track the locally issued hurricane evacuation notices. The text data were collected mainly with a spatially targeted web scraping method. They were manually labeled and then classified using natural language processing techniques with deep learning models. The classification of mandatory evacuation notices achieved a high accuracy (recall = 96%). We used Hurricane Ian (2022) to illustrate how real-time evacuation notices extracted from local government sources could be redistributed with a Web GIS system. Our method applied to future hurricanes provides live data for situation awareness to higher-level government agencies and news media. The archived data helps scholars to study government responses toward weather warnings and individual behaviors influenced by evacuation history. The framework may be applied to other types of disasters for rapid and targeted retrieval, classification, redistribution, and archiving of real-time government orders and notifications.

# 1. Introduction

In the event of disasters such as hurricanes, local authorities are responsible for issuing evacuation notices in the United States (Cigler 2009, Kruger et al. 2020). These governing bodies typically operate independently to distribute their evacuation notices through their own channels (such as press releases, alert systems, websites, and social media) as well as local TV and news outlets (Veil, Buehner and Palenchar 2011). Unlike weather warnings, there is currently no centralized platform dedicated to providing real-time evacuation notices nationwide. Consequently, when it comes to large-scale evacuations (Lindell et al. 2018a), it is not uncommon for higher-level government agencies, news outlets, or other concerned organizations to encounter difficulties in effectively tracking real-time orders in local communities.

We believe a centralized web portal, which distributes and stores real-time hurricane evacuation notices across different government agencies throughout the United States, is beneficial to several stakeholder groups. First, the regional and state emergency managers and government officials may visit this web portal to maintain situational awareness of the ongoing evacuation situation inside and beyond their political boundaries. Second, emergency response groups as well as federal emergency officials may have a better understanding of where their services and support might be most needed. Third, the public may look for active evacuation notices directly through this web portal. News outlets may also redistribute real-time evacuation notices collected by this web portal to those who do not have direct access to it. Lastly, the archived evacuation notice data are useful to scholars, especially those interested in understanding (1) patterns between hurricane forecasting and local emergency-response practices, (2) phases and timing of local emergency management practices and their potential socioeconomic consequences, and (3) impacts of evacuation history ("cry wolf" cases) on individual citizen's evacuation decision-making.

To create this web portal, we developed a workflow for the automated extraction and classification of real-time hurricane evacuation notices (Figure 1). Evacuation notices for the eastern United States since 2018 originated mainly from government officials' social media and websites. We collected the related text data primarily using web scraping, aided with a spatial filter so as to target the governments' official information channels. The earlier text data between 2001 and 2017 were collected manually from multiple Internet sources that included official government websites but not social media, which were too dated and hence not publicly available online. We manually labeled these text data. We first distinguished "Evacuation Notice" from "Not an Evacuation Notice". For an Evacuation Notice, we further labeled "Mandatory Evacuation Notice" and "Voluntary Evacuation Notice" separately. We applied the latest Natural Language Processing (NLP) techniques to classify all text data into different categories. Particular attention was given to classifying mandatory evacuation notices with a high recall measure, for which the value of 1.0 indicates that 100% of the actual mandatory evacuation notices are correctly identified using our Deep Learning (DL) models. Retrieved and classified data can then be redistributed promptly to different stakeholders to facilitate more effective disaster responses.

The major contributions of our work are as follows: (1) We included geographic context data for rapid and precise identification of an emerging evacuation situation through governments' official social media. (2) The DL models extracted evacuation notices from text data with high accuracies. (3) The classified text data, when published on Web GIS systems, provide real-time evacuation notices for government officials, scholars, and the general public. And (4) the data are archived for future use by scholars and other interested stakeholders. They may be integrated into the current federal or state emergency management or communication systems. The future applications may include rapid collection and dissemination of additional real-time evacuation information (such as shelter, route, and other resources) from official, reliable information channels for hurricanes and other types of disasters.

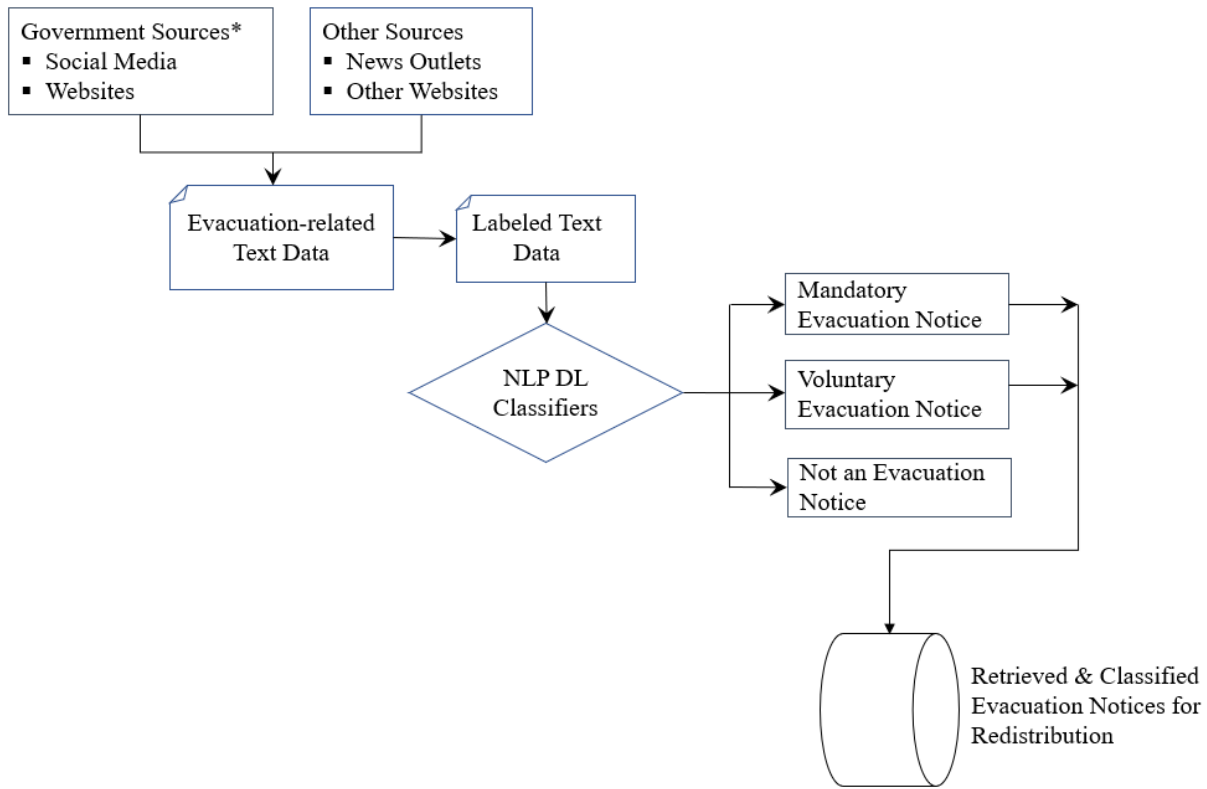

Figure 1. Workflow of evacuation notice information extraction and classification. NLP: Natural Language Processing. DL: Deep Learning. *Government social media and websites were targeted with a spatial filter generated based on geographic identifiers[1] in National Weather Service alerts.

## 2. Research Background
While general guides on hurricane evacuations are available through multiple government and organization websites (e.g., https://www.ready.gov/evacuation), the centralized web portal that is specialized in providing real-time evacuation notices throughout the United States is absent. Many state governments have their own websites to disseminate hurricane-related information. For example, Texas has a hurricane center[2], Louisiana has an emergency office[3], and Florida has a division of emergency management[4]. However, none of these state-level emergency management offices provide real-time evacuation notices except the Florida website[5] that provided some of the active hurricane evacuation notices issued within the State of Florida.

---

[1] https://www.weather.gov/media/documentation/docs/NWS_Geolocation.pdf
[2] https://gov.texas.gov/hurricane
[3] https://gohsep.la.gov/emergency/
[4] https://www.floridadisaster.org/hazards/hurricanes/

*2.1. Challenges of Building the Database of Real-Time Evacuation Notices*
In the United States, it is difficult for a single government agency or organization to track hurricane evacuation notices without a good deal of manpower or advanced AI techniques. The reasons are as follows: First, hurricane evacuation notices may be issued across a large geographic area. Hurricane Floyd (1999), originally forecasted to strike Florida, moved parallel to the east coast of the United States and caused mass evacuations in Florida, Georgia, South Carolina, North Carolina, and Virginia (Dow and Cutter 2002). Hurricane Gustav (2008), predicted to be a stronger hurricane than the actual intensity at landfall in Louisiana, triggered mandatory and voluntary evacuations in Louisiana, Texas, and Alabama (Senkbeil et al. 2010). Hurricane Irene (2011), with a projected path over much of the U.S. east coast, resulted in mass evacuations in North Carolina, Virginia, Maryland, Delaware, New Jersey, and low-lying areas of New York City (Ng, Diaz and Behr 2016). Hurricane Irma (2017) generated one of the largest evacuations in recent U.S. history along the Florida, Georgia, and South Carolina coasts (Martín, Cutter and Li 2020).

Second, hurricane evacuation notices are issued and distributed by hundreds of individual governing bodies at different administrative levels (such as state, county, and city/town). In the United States, evacuation notices are issued by state or local governments in disastrous situations (Connolly, Klofstad and Uscinski 2020). In states such as Georgia and South Carolina, the state governor issues hurricane evacuation notices. However, evacuation notices are more commonly issued by local governments, such as counties or parishes (in Louisiana). In some New England states (such as Connecticut and Rhode Island) and Virginia's administrative units, evacuation notices are issued by cities or towns. These government agencies are also responsible for notifying concerned citizens of an effective evacuation notice.

Third, evacuation notices change dynamically during the development of a hurricane. They are often issued and updated following changes in the predicted track or intensity of the hurricane (Baker 2000, Regnier 2008). For example, according to the real-time data we collected for Hurricane Dorian (2019), this storm prompted coastal counties in Florida to issue evacuation notices, mostly between August 30 and September 1. Georgia and South Carolina governors issued evacuation notices on September 1. The North Carolina governor issued evacuation notices on September 3. Governing bodies in Virginia issued evacuation notices on September 5. In Florida, a few counties (such as Osceola, Saint Lucie, and Flagler) issued a voluntary evacuation notice on August 30 and updated it to a mandatory notice on September 1 or September 2. The large-scale, dynamic changes in official-issued evacuation notices have also been observed for the manually collected historical data, such as with Hurricane Irma in 2017 (Figure 2).

Lastly, the government's websites and social media provide reliable channels for the spread of evacuation notices (Graham, Avery and Park 2015, Knox 2022). Meanwhile, these official channels also disseminate other protective action information such as storm watches and warnings, evacuation route advisories, and resource availability such as sandbags and public shelter information (Wukich 2016, Shi 2021). From the technical aspect of real-time data collection, the high velocity, limited veracity, and large volume of social media data make it difficult to identify relevant information pertaining to a specific disaster-related theme or category, such as hurricane evacuation notices (Liu et al. 2019a, Naaz, Ul Abedin and Rizvi 2021).

---

[5] https://www.floridadisaster.org/evacuation-orders/

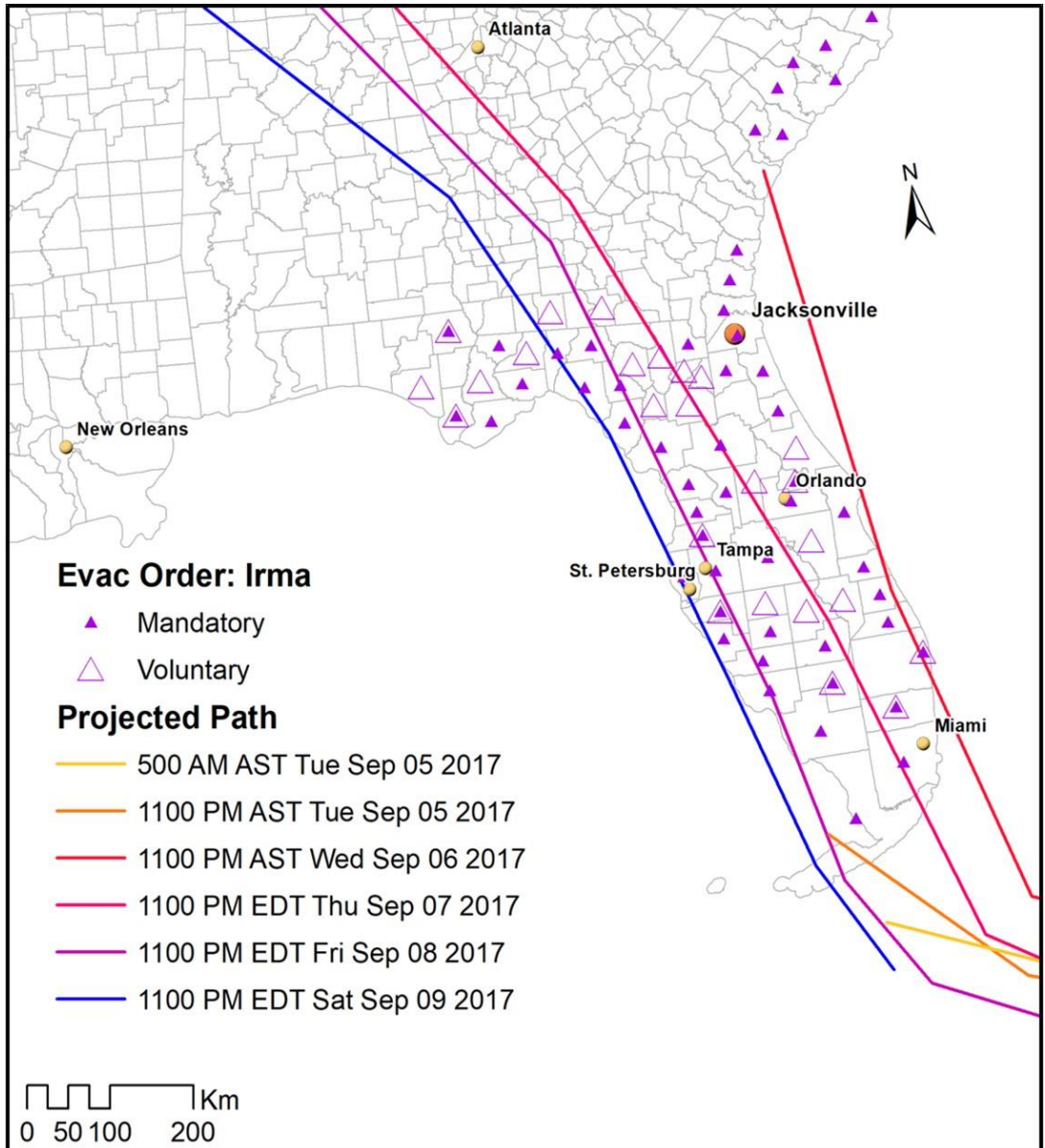

Figure 2. Evacuation notices made by the officials in response to Hurricane Irma (2017) along with the updated predictions of storm track.

## *2.2. Opportunities for Building the Database of Real-Time Evacuation Notices*

The increasing dissemination of evacuation notices online makes it feasible to detect hurricane evacuation orders through the Internet. Social media, such as Twitter and Facebook, play an important role in spreading disaster-related information in the emergency planning and response stages (Houston et al. 2014). They also facilitate the rapid distribution of weather warnings, evacuation notices, the need for help or volunteers, and other types of time-critical information (Alam, Ofli and Imran 2020).

The maturing web scraping techniques enable the retrieval of hurricane evacuation orders from the Internet in a timely manner. Web scraping is a computerized method to obtain a large volume of information from the Internet (Zhao 2017, Zhou, Wang and Li 2017, Tan and Guan 2021, Bale et al. 2022). In disaster-related research, this includes the process of automatically searching social media to rapidly detect real-time situations in different disaster phases and maintain situational awareness (Huang and Xiao 2015, Wukich 2016, Ceccato and Petersson 2022). Previous work utilized specific hashtags, semantic filtering, and keyword-based filtering to subset social media data and extract relevant hazard-related posts (Mohanty et al. 2021). Location and time of events have also been used to narrow down tweets for event detection and situational awareness (Spinsanti and Ostermann 2013, de Albuquerque et al. 2015, Farnaghi, Ghaemi and Mansourian 2020, Scheele, Yu and Huang 2021).

The improved machine learning techniques make it possible to classify the scraped, unstructured text data automatically with pre-trained algorithms (Hu, Deng and Zhou 2019, Jiang, Li and Cutter 2019, Martin and Schuurman 2020, He et al. 2023). Modern natural language processing (NLP) methods can be used to extract and classify disaster-related notifications. However, they have not been applied to targeting evacuation notices in previous hazard-related studies. As shown in recent review articles about social media data extraction for situational awareness during natural hazards (Granell and Ostermann 2016, Vongkusolkit and Huang 2021), little emphasis was put on the content classification of official notices or notifications in the disaster classification schema of social media data. Many of the studies generated keywords such as "issued" (Sit, Koylu and Demir 2019), "warning" (Wang and Stewart 2015), "evacuees" (Alam et al. 2020), and "evacuate" (Wang and Ye 2019); but none of the studies were targeted specifically on evacuation notices. In addition, the classification of disaster-related text data still has a lot of room for improvement. A study retrieving geo-tagged Twitter data for Hurricane Sandy classified tweets into different categories including "evacuation" (which was used in a much broader context than just evacuation notices) and achieved an accuracy of 64% for this category (Huang and Xiao 2015).

We believe an application of NLP that uses deep learning (DL) methods can improve the classification accuracy of the scraped, unstructured evacuation-related text. DL is a type of machine learning method based on artificial neural networks that attempt to simulate the human brain to learn from data (Grekousis 2019). It uses multiple layers to progressively extract higher-level features from the input data (Li, Hsu and Hu 2021, Liao et al. 2023). DL has been the state-of-the-art method for modern NLP research and applications (Acheampong, Nunoo-Mensah and Chen 2021). For example, a convolutional neural network (CNN) was shown to be a superior technique to logistic regression or support vector machine for classifying tweets into situational-awareness categories for hurricane-related events (Yu et al. 2019). DL methods for emotion classification and behavior modeling were also used to extract true emergency tweets (Dwarakanath et al. 2021, Hembree et al. 2021). Most recently BERT-based models with a customized CNN classifier were used to categorize rescue-related tweets with high accuracies (Zhou et al. 2022). BERT stands for Bidirectional Encoder Representations from Transformers.

In this study, we selected the BERT-based DL methods, i.e., BERT (Devlin et al. 2019) and RoBERTa (Liu et al. 2019b), for our classification of evacuation-related text data. BERT pre-trains deep bidirectional representation from the unlabeled text by jointly conditioning on both the left and right context. It is pre-trained using masked language modeling and next-sentence prediction (Devlin et al. 2019). It uses transformers (Vaswani et al. 2017) as its basic architecture, which effectively extracts the relational context of the text, i.e., the notice of words and their relationships in a sentence. RoBERTa, which stands for Robustly optimized BERT approach, has the same architecture as BERT but is pre-trained using dynamic masking and full-sentence sample without the next-sentence prediction. In addition, its text database for pre-training is ten times the size of the BERT text database (Liu et al. 2019b).

## 3. Methods

### 3.1. Data Collection and Labeling

We collected data for the current study by two means. For more recent data (2018 and later), we searched county and state governments' tweets and Facebook posts containing the keywords "hurricane" and "evacuate" (including its variants such as evacuation, evacuated, and so on) using three Python packages twitter-scraper[6], facebook-scraper[7] and Beautiful Soup[8]. We targeted counties that received hurricane warnings, hurricane watches, tropical storm warnings, tropical storm watches, or tropical cyclone statements from the National Weather Service (NWS) with NWS APIs[9]. We used the NWS APIs to identify warning/watch-associated geocode that returns county Federal Information Processing Standard (FIPS) codes. The FIPS codes were then used to match records in our location spreadsheet, which included the county's name, the state that the county belongs to, the county's government website, its emergency management website, and its social media such as Twitter and Facebook, etc. For data since 2018, we also manually collected evacuation notices through notifications, news, announcements, and alerts published through government websites and news outlets, so as not to miss any evacuation notices if any governing bodies do not use social media as their posting channel. The earlier data from 2001 to 2017, which were no longer accessible through social media, were collected manually through government websites, news websites, and other reports on the Internet.

After the relevant text data were collected, three annotators were trained to label all text collected from 2001 to 2020. They labeled each text as an evacuation notice or non-evacuation notice. For each evacuation notice, the annotators labeled it as a mandatory notice or voluntary notice. Guidelines for the labeling were provided and the annotators were trained with the same set of sample data. Two annotators worked independently and the third checked their work. When discrepancy occurred, the third annotator worked directly with the leading researcher to decide on the final labeling.

### 3.2. Text Data Classification

The labeled dataset was then split into training, validation, and testing subsets to classify text data into different categories. We examined two different classification approaches. First, we classified each record by whether it was an evacuation notice, i.e., Evacuation Notice vs. Not an Evacuation Notice. Second, we classified all records into three categories including two specific types of evacuation notices, i.e., Mandatory Evacuation Notice, Voluntary Evacuation Notice, or Not an Evacuation Notice (Figure 1).

---

[6] https://github.com/bisguzar/twitter-scraper
[7] https://github.com/kevinzg/facebook-scraper
[8] https://www.crummy.com/software/BeautifulSoup/
[9] https://www.weather.gov/documentation/services-web-api

The DL models, i.e., BERT (Devlin et al. 2019) and RoBERTa (Liu et al. 2019b), were applied to classify those texts in the Python package of Transformers (Wolf et al. 2020). We used a DL architecture with a pre-trained transformer-based language model as an encoder for our classification tasks in this study (Figure 3). Each text was split into tokens using the WordPiece (Wu et al. 2016) tokenizer and a special token [CLS] was added at the beginning of the text. These tokens went through the BERT or RoBERTa model for encoding to output a text representation for final classification. We used the embedding output of the [CLS] token as the text representation and fed it into a linear layer. The output of the linear layer was then fed into the softmax classifier to classify the texts.

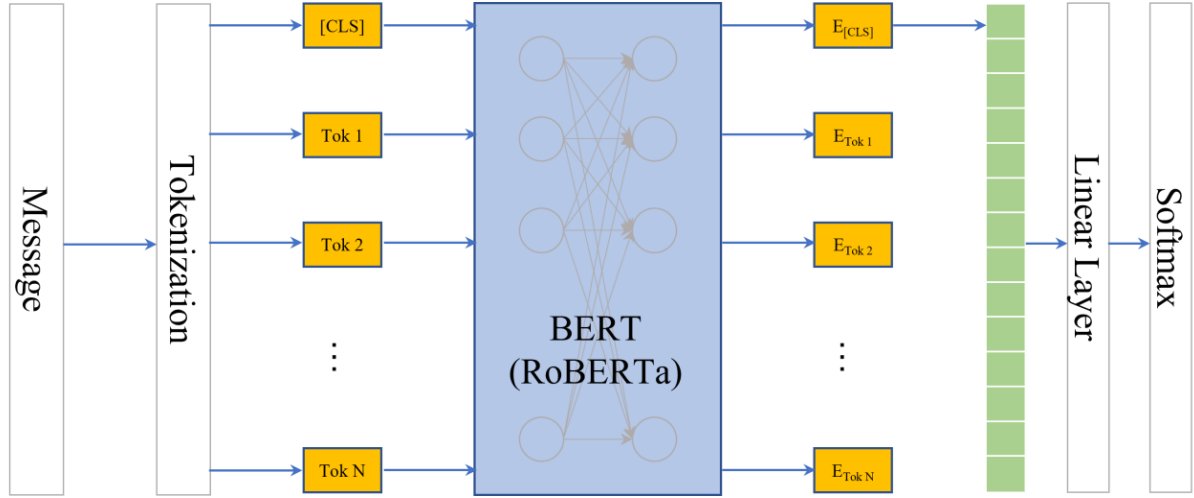

Figure 3. Architecture of the classification models

In our experiments, we set the maximal sequence length at 512 tokens in accordance with the maximal length of the text. We used a batch size of 4 due to the constraint of the GPU memory. We trained the models using a CrossEntropy loss function with a learning rate of 5.0e-06 and AdamW as the optimizer. We used an early stop to pick the model with the best performance on the validation data during the training process. We evaluated the performance of the models using metrics of precision, recall, F1 score, and accuracy. The 10-fold cross-validation was used by splitting the data into 10 folds with 8 folds for training, 1 fold for validation, and 1 fold for testing in each run.

## 4. Results
### *4.1. Retrieved Data from the Internet and Data Labeling*
A total of 1,305 texts published by 218 counties in 19 states since 2001 were collected (Figure 4). When the same evacuation notice was encountered more than once as a result of the government using different wording in multiple posts, these texts were considered to be non-duplicates from the language processing point of view. The number of evacuation texts varied greatly each year (Figure 5), with the most abundant texts in 2012 (Hurricane Isaac and Hurricane Sandy), 2017 (Hurricane Harvey and Hurricane Irma), 2005 (Hurricane Dennis and Hurricane Katrina), and 2020 (multiple hurricanes especially for State of Louisiana).

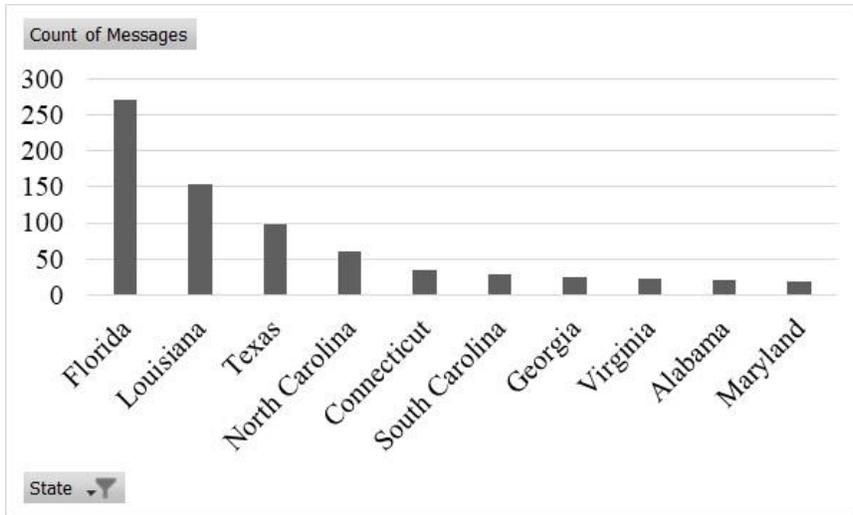

Figure 4. Count of evacuation-related texts by state. Only top ten states are shown.

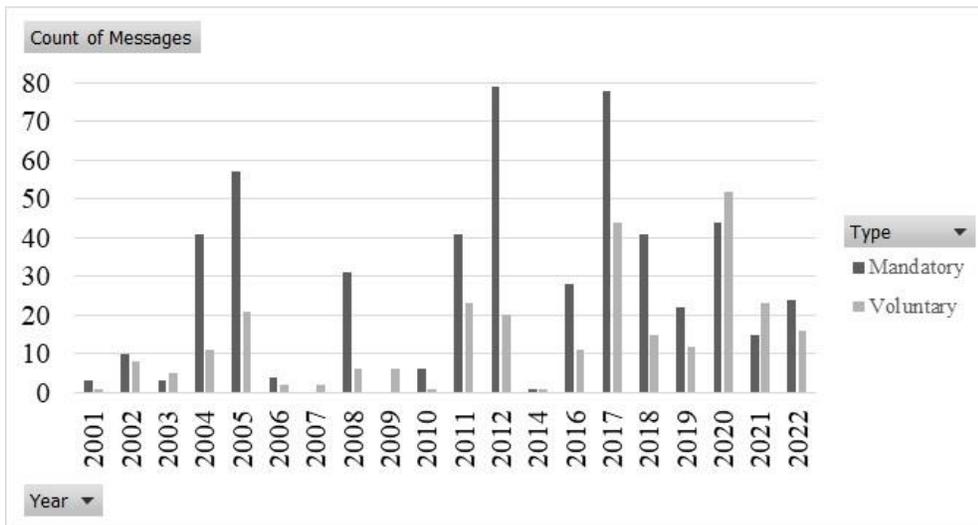

Figure 5. Numbers of mandatory and voluntary evacuation notices by year

The annotators labeled 1,227 texts, which included 578 (47%) of the texts collected from Twitter and Facebook and 649 (53%) of the texts collected from county government websites or news outlets for the 2001-2020 time period. For government Twitter and Facebook posts, 127 of the 578 retrieved texts were evacuation notices. Among the 1,227 labeled texts, 730 (59%) of the texts were labeled as evacuation notices, and 497 (41%) of the texts were labeled as non-evacuation notices. The evacuation notices contained 489 (67%) mandatory notices and 241 (33%) voluntary notices.

### *4.2. Deep Learning Classification of the Evacuation Notice Data*
We experimented using BERT and RoBERTa as encoders in the classification models. Both models performed consistently with small variations when we classified all texts into two categories, i.e., Evacuation Notice vs. Not an Evacuation Notice. For the classification of whether a retrieved text was an evacuation notice, the model using BERT as encoders achieved a weighted F1 score and accuracy of approximately 91% and the model using RoBERTa achieved 92% (Table 1). The accuracy measure indicates correctly predicted cases in the total cases, with 1.0 as 100% of the cases being correctly classified. Both models correctly classified 91-92% of the evacuation-related texts.

Table 1. Performance for the classification of whether a text is an evacuation notice based on 10-fold cross-validation. In total, 1,227 evacuation-related texts were classified. These included 730 texts labeled as "evacuation notice" by researchers.

|  | Uncased BERT Base Model | | | RoBERTa Base Model | | |
| --- | --- | --- | --- | --- | --- | --- |
|  | Precision | Recall | F1 | Precision | Recall | F1 |
| Evacuation Notice | 0.885 (0.052) | 0.975 (0.017) | 0.927 (0.029) | 0.897 (0.039) | 0.978 (0.017) | 0.935 (0.023) |
| Not an Evacuation Notice | 0.958 (0.028) | 0.810 (0.095) | 0.874 (0.058) | 0.963 (0.030) | 0.834 (0.067) | 0.892 (0.043) |
| Macro Avg | 0.922 (0.030) | 0.892 (0.047) | 0.901 (0.043) | 0.930 (0.026) | 0.906 (0.035) | 0.914 (0.033) |
| Weighted Avg | 0.915 (0.033) | 0.908 (0.038) | 0.906 (0.040) | 0.924 (0.028) | 0.919 (0.030) | 0.918 (0.031) |
| Accuracy |  |  | 0.908 (0.038) |  |  | 0.919 (0.030) |

The highest possible value of an F-score is 1.0, indicating perfect precision and recall. Precision for "evacuation notice" refers to the percentage of true notices to the total of the predicted evacuation notices (which include both correctly and wrongly predicted notices). In other words, 88.5% of our model-predicted evacuation notices were truly evacuation notices when using BERT and 89.7% were truly evacuation notices when using RoBERTa (Table 1). Recall for "evacuation notice" refers to the ratio of the percentage of correctly predicted evacuation notices to the total number of true evacuation notices. Thus, our models correctly identified over 97% (97.5% for BERT and 97.8% for RoBERTa) of the actual evacuation notices in this dataset (Table 1).

The two models using BERT or RoBERTa also classified evacuation notices by types (mandatory vs. voluntary) with high accuracies, with F1 and accuracy at 87.6% for BERT and 88.2% for RoBERTa (Table 2). Both models correctly classified approximately 88% of evacuation-related texts into their correct types. For the "mandatory evacuation notice" category, the F1 score was approximately 89%. When BERT was used, approximately 83% (precision) of the model-predicted mandatory notices were true mandatory evacuation notices and approximately 96% (recall) of the true mandatory notices were identified through our models. When RoBERTa was used, recall declined (93.7%) while precision increased (85.8%). For the "voluntary evacuation notice" category, the F1 score was above 82%. When BERT was used, approximately 84% (precision) of the model-predicted voluntary notices were true voluntary notices and approximately 85% (recall) of the true voluntary notices were identified through our models. When RoBERTa was used, recall increased slightly (85.0%) while precision decreased (79.9%).

Table 2. Performance of the classification by types of evacuation notices based on 10-fold cross-validation. In total, 1,227 evacuation-related texts were classified. These included 489 texts labeled as "mandatory evacuation notice", 241 texts labeled as "voluntary evacuation notice", and 497 texts labeled as "not an evacuation notice" by research annotators.

|  | Uncased BERT Base Model | | | RoBERTa Base Model | | |
| --- | --- | --- | --- | --- | --- | --- |
|  | Precision | Recall | F1 | Precision | Recall | F1 |
| Mandatory Evacuation Notice | 0.829 (0.056) | 0.961 (0.027) | 0.889 (0.036) | 0.858 (0.048) | 0.937 (0.027) | 0.895 (0.030) |
| Voluntary Evacuation Notice | 0.835 (0.074) | 0.846 (0.090) | 0.838 (0.066) | 0.799 (0.083) | 0.850 (0.071) | 0.822 (0.069) |
| Not an Evacuation Notice | 0.972 (0.032) | 0.804 (0.067) | 0.878 (0.044) | 0.961 (0.032) | 0.841 (0.066) | 0.896 (0.043) |
| Macro Avg | 0.879 (0.034) | 0.870 (0.047) | 0.868 (0.043) | 0.873 (0.042) | 0.876 (0.042) | 0.871 (0.042) |
| Weighted Avg | 0.889 (0.030) | 0.876 (0.038) | 0.876 (0.039) | 0.890 (0.034) | 0.882 (0.036) | 0.882 (0.036) |
| Accuracy |  |  | 0.876 (0.038) |  |  | 0.882 (0.036) |

## 5. Discussion

### 5.1. Retrieval of the Evacuation Notice Data from the Government Official Channels

In our search of government Twitter and Facebook posts, 127 of 578 retrieved texts were evacuation notices. This proportion (22%) was achieved after we specifically targeted the potentially affected geographic areas as indicated by the hurricane/tropical storm watches and warnings distributed by NWS for a given hurricane. This procedure takes place in real-time and helps us to quickly identify counties that have issued evacuation notices for an approaching hurricane. The knowledge of the government's official channels also provides a much more efficient approach to detecting the original, official evacuation notices, compared to filtering out the evacuation notices from the social media data flooded by public users' posts on hurricane- or non-hurricane-related matters.

A database of historical hurricane evacuation notices is currently absent in the United States. Nor is it easy to obtain records of historical evacuation notices from local emergency offices since they are not required to archive these data according to personal communication with hurricane researchers and county officials. We manually searched government, news, and other websites including Internet Archive (https://archive.org/). Historical evacuation notices issued before 2000 are less likely to be harvested using this approach. This indicates the need to build such a database for the future knowledge of evacuation history for stakeholders such as scholars, government agencies and the general public.

*5.2. Deep Learning Methods and Classification Accuracies*

The differences between the performances of BERT and RoBERTa were not very significant in our study. Both models yielded very high measures of recall, i.e., 97%, for evacuation notices regardless of the type. For research implementation, we favor the three-category classification (Figure 1) because knowing the specific types of evacuation notices, i.e., mandatory or voluntary, has practical implications for individuals' and communities' decision-making and preparation for the approaching hurricane. BERT achieved a recall of 96% for mandatory evacuation notices; thus, this model can be used to classify the scraped, unstructured, real-time evacuation-related text data in future hurricane seasons. The measure of recall is very important for our research since we do not want to miss any of the real mandatory evacuation notices.

Challenges existed for the labeling and classification of hurricane evacuation-related texts. The following discussion focuses on mandatory evacuation notices mainly using the original text as examples. All the case numbers cited hereafter are listed in Appendix A. Many text data extracted from the government's websites or social media include both mandatory and voluntary notices (e.g., case 175, 1283). Sometimes they include the update of a voluntary order to a mandatory order (e.g., case 175) or the expansion of evacuation by zones/area/population groups. Text data can be wordy and may include extra information (such as storm conditions, curfew information, and evacuation routes) other than evacuation notices (e.g., case 528, 1214). The text message wording can be untypical; for example, case 1214 used "I" to address the authority who issued the evacuation notices, whereas in most cases a governing body or its representative is specified (i.e., all other cases in Appendix A). The tense of verbs may also influence the algorithms' classification accuracy such as the case where an order "will be issued" was not counted as a mandatory evacuation notice (case 725, 1283).

*5.3. Redistributing the Retrieved and Classified Real-Time Evacuation Notices*

In the United States, state and local governments have the authority to issue and distribute hurricane evacuation notices. Previous studies documented people's preference for using local government as their evacuation information sources (DeYoung et al. 2016) as well as the benefits of using official sources, e.g. the increased consistency of people's perceived information (Lee et al. 2021). However, the effectiveness of the government's websites and social media in disseminating real-time evacuation notices directly to the public remains uncertain. Recent studies documented older and less-educated populations are underserved through social media by local agencies in hazard communications (Neely and Collins 2018). In the data collection process for this study, we also observed service disruptions when using the government's official websites due possibly to the high traffic. For example, before the landfall of Hurricane Ian, Sarasota County in Florida experienced a website outage on September 26, 2022. Lee County, Florida put a Facebook post out on September 28, 2022, to warn its citizens about the potential instability of the county's website that may disrupt viewers for hurricane-related information (including evacuation notices). These indicate the need to have the evacuation notices hosted on other virtual spaces than the government's official websites alone.

The outcomes of our retrieved and classified real-time evacuation data may be transposed into a web mapping system that is hosted on a scalable cloud infrastructure (such as ESRI Managed Cloud Services or Amazon Web Services) outside of the hurricane risk areas to allow for stable public access. We published evacuation notices for Hurricane Ian (2022) on ESRI's ArcGIS Online (Figure 6). It was made using the ArcGIS Web AppBuilder and is accessible to the general public. Users can visually check the locations of evacuation notices in effect. They can input an address or place name to see whether and what type of evacuation notice is in effect in that area. The actual text of evacuation notices (such as who is affected and, if applicable, when the notice becomes effective) and the official government source of the evacuation notice are all provided.

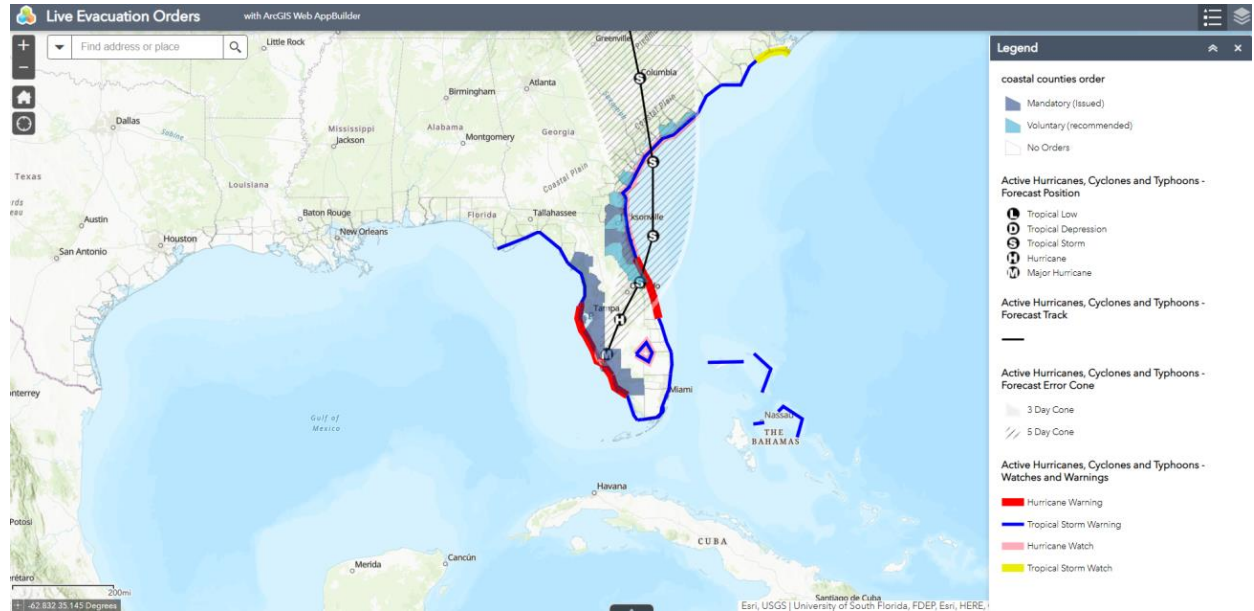
Figure 6. Real-time evacuation notices for Hurricane Ian, 2022

The web interface with real-time evacuation notices helps cross-division government agencies and the public to understand the geographic range of the affected areas. This may facilitate coordination among different audiences to reduce hurricane-related risks (Miyazaki, Nagai and Shibasaki 2015). For example, the rapid and accurate communication of evacuation notices helps intergovernmental coordination of transportation, shelter, and other resources for the approaching emergency (Carpender et al. 2006, Baker 2017). It also helps the affected individuals to make evacuation decisions (Baker 1991, Lindell et al. 2018b). The web interface also incorporates other critical hurricane-related information such as hurricane warning information and the active hurricane forecast from the National Hurricane Center (NHC).

### 5.4. Research Limitations and Future Work
In the real-time data collection process, we encountered some difficulties with the inclusion of all hurricane evacuation notices when we relied on the government's official social media alone. Some of the local governments appeared not to distribute their evacuation notices through social media. For this reported work, we manually collected evacuation notices through their county website, their emergency management office's website, and occasionally from local news outlets. This challenge points out the need for future research on using advanced text mining techniques as well as improved hazard ontology (Wang and Stewart 2015, Huang, Cervone and Zhang 2017, Hu 2018) to collect Internet data on those less structured web pages (such as county or news media websites) compared to Twitter or Facebook.

For the classification of evacuation-related text data, our results meant 3-4% of the actual evacuation notices are missing with the current NLP DL classification. This may be corrected through additional model training to target a higher recall, which may adversely decrease the precision. Alternative approaches include: (1) inviting the local governing bodies to check our evacuation notice posts based on the NLP classification and (2) through the Web GIS interface we may collect user feedback and correct miss-classified evacuation notices. Future research shall also incorporate more training data and classify them to include more structured outcomes (such as who is expected to evacuate, the time at which the notice is effective, and so on) with advanced named-entity recognition (NER) approaches. In addition, the same evacuation notice that was posted with different wording needs to be consolidated especially for users to whom the count of non-duplicating evacuation notices is necessary.

In the future, we will compile and distribute the retrieved and classified real-time evacuation notices with ArcGIS API for JavaScript, which allows for increased automation and flexibility for data management and sharing. User inputs on whether an evacuation notice was correctly provided in our web interface can be collected and used to correct and train our classification models. This web interface also has the potential for data integration and exchange with the existing federal or state emergency management software. Thus, it supports the interactive communication of emergency information between federal and local agencies, among local agencies, and between individuals and government agencies.

## 6. Conclusions

In this study, we used the automated retrieval and classification of hurricane evacuation notices as an example to illustrate how modern NLP techniques (e.g., BERT) and a centralized web system can substantially increase the findability and accessibility of the real-time, officially released disaster announcements that are isolated and scattered throughout the Internet. As a result, we constructed a database of hurricane evacuation notices in the US since 2001. Using our targeted web scraping approach, evacuation notices were collected mainly from reliable government sources since 2018. With the state-of-the-art NLP DL methods, the classification accuracy (recall) of mandatory evacuation notices was above 96%.

This method will be used to collect, classify, disseminate, and archive real-time evacuation notices issued through numerous local government agencies for hurricanes in the future. The real-time data may be used by higher-level government agencies for situation awareness and by news media for information redistribution to the general public. The archived evacuation notice data are valuable for hazard research communities to study government responses toward weather warnings as well as individual evacuation behaviors related to the past experience of evacuation advisories. The framework may also be applied to other types of natural and manmade disasters, where rapid and targeted retrieval and redistribution of loosely distributed, unstructured information from a large number of government entities is desired.

**References**

Acheampong, F. A., H. Nunoo-Mensah & W. Y. Chen (2021) Transformer models for text-based emotion detection: a review of BERT-based approaches. *Artificial Intelligence Review*.

Alam, F., F. Ofli & M. Imran (2020) Descriptive and visual summaries of disaster events using artificial intelligence techniques: case studies of Hurricanes Harvey, Irma, and Maria. *Behaviour & Information Technology,* 39**,** 288-318.

Baker, E. J. (1991) Hurricane evacuation behavior. *International Journal of Mass Emergencies and Disasters,* 9**,** 287-310.

---. 2000. Hurricane evacuation in the United States. In *Storms,* eds. P. R. Jr. & P. RA., 306–319. New York: London: Routledge.

Baker, K. (2017) Reflection on Lessons Learned: An Analysis of the Adverse Outcomes Observed During the Hurricane Rita Evacuation. *Disaster Medicine and Public Health Preparedness,* 12**,** 115 - 120.

Bale, A. S., N. Ghorpade, S. R, S. Kamalesh, R. R & S. R. B. 2022. Web Scraping Approaches and their Performance on Modern Websites. In *2022 3rd International Conference on Electronics and Sustainable Communication Systems (ICESC)*, 956-959.

Carpender, S. K., P. H. Campbell, B. J. Quiram, J. C. Frances & J. J. Artzberger (2006) Urban Evacuations and Rural America: Lessons Learned from Hurricane Rita. *Public Health Reports,* 121**,** 775 - 779.

Ceccato, V. & R. Petersson (2022) Social Media and Emergency Services: Information Sharing about Cases of Missing Persons in Rural Sweden. *Annals of the American Association of Geographers,* 112**,** 266-285.

Cigler, B. A. (2009) Post-Katrina Hazard Mitigation on the Gulf Coast. *Public Organization Review,* 9**,** 325-341.

Connolly, J. M., C. A. Klofstad & J. E. Uscinski. 2020. Leaving home ain't easy: Citizen compliance with local government hurricane evacuation orders.

de Albuquerque, J. P., B. Herfort, A. Brenning & A. Zipf (2015) A geographic approach for combining social media and authoritative data towards identifying useful information for disaster management. *International Journal Of Geographical Information Science,* 29**,** 667-689.

Devlin, J., M.-W. Chang, K. Lee & K. Toutanova. 2019. BERT: Pre-training of Deep Bidirectional Transformers for Language Understanding. In *NAACL*.

DeYoung, S. E., T. Wachtendorf, A. K. Farmer & S. Penta (2016) NOAA Radios and Neighbourhood Networks: Demographic Factors for Channel Preference for Hurricane Evacuation Information. *Political Behavior: Cognition*.

Dow, K. & S. L. Cutter (2002) Emerging Hurricane Evacuation Issues: Hurricane Floyd and South Carolina. *Natural Hazards Review,* 3**,** 12-18.

Dwarakanath, L., A. Kamsin, R. A. Rasheed, A. Anandhan & L. Shuib (2021) Automated Machine Learning Approaches for Emergency Response and Coordination via Social Media in the Aftermath of a Disaster: A Review. *Ieee Access,* 9**,** 68917-68931.

Farnaghi, M., Z. Ghaemi & A. Mansourian (2020) Dynamic Spatio-Temporal Tweet Mining for Event Detection: A Case Study of Hurricane Florence. *International Journal of Disaster Risk Science,* 11**,** 378-393.

Graham, M. W., E. J. Avery & S. Park (2015) The role of social media in local government crisis communications. *Public Relations Review,* 41**,** 386-394.

Granell, C. & F. O. Ostermann (2016) Beyond data collection: Objectives and methods of research using VGI and geo-social media for disaster management. *Computers, Environment and Urban Systems,* 59**,** 231-243.

Grekousis, G. (2019) Artificial neural networks and deep learning in urban geography: A systematic review and meta-analysis. *Computers, Environment and Urban Systems,* 74**,** 244-256.

He, H., R. Sun, J. Li & W. Li (2023) Urban landscape and climate affect residents' sentiments based on big data. *Applied Geography,* 152**,** 102902.

Hembree, A., A. Beggs, T. Marshall & E. N. Ceesay. 2021. Decoding Linguistic Ambiguity in Times of Emergency based on Twitter Disaster Datasets. In *IEEE 11th Annual Computing and Communication Workshop and Conference (CCWC)*, 1239-1244. Electr Network: Ieee.

Houston, J. B., J. Hawthorne, M. Perreault, E. Park, M. Hode, M. Halliwell, S. McGowen, R. Davis, S. Vaid, J. McElderry & S. Griffith (2014) Social media and disasters: A functional framework for social media use in disaster planning, response, and research. *Disasters,* 39.

Hu, Y., C. Deng & Z. Zhou (2019) A Semantic and Sentiment Analysis on Online Neighborhood Reviews for Understanding the Perceptions of People toward Their Living Environments. *Annals of the American Association of Geographers,* 109**,** 1052-1073.

Hu, Y. J. (2018) Geo-text data and data-driven geospatial semantics. *Geography Compass,* 12.

Huang, Q., G. Cervone & G. Zhang (2017) A cloud-enabled automatic disaster analysis system of multi-sourced data streams: An example synthesizing social media, remote sensing and Wikipedia data. *Computers, Environment and Urban Systems,* 66**,** 23-37.

Huang, Q. & Y. Xiao (2015) Geographic Situational Awareness: Mining Tweets for Disaster Preparedness, Emergency Response, Impact, and Recovery. *Isprs International Journal of Geo-Information,* 4**,** 1549-1568.

Jiang, Y., Z. Li & S. L. Cutter (2019) Social Network, Activity Space, Sentiment, and Evacuation: What Can Social Media Tell Us? *Annals of the American Association of Geographers,* 109**,** 1795-1810.

Knox, C. C. (2022) Local emergency management's use of social media during disasters: case study of Hurricane Irma. *Disasters*.

Kruger, J., M. J. Smith, B. Chen, B. Paetznick, B. M. Bradley, R. Abraha, M. Logan, E. R. Chang, G. Sunshine & S. Romero-Steiner (2020) Hurricane Evacuation Laws in Eight Southern U.S. Coastal States — December 2018. *Morbidity and Mortality Weekly Report,* 69**,** 1233 - 1237.

Lee, S., B. C. Benedict, Y. G. Ge, P. M. Murray-Tuite & S. V. Ukkusuri (2021) An application of media and network multiplexity theory to the structure and perceptions of information environments in hurricane evacuation. *Journal of the Association for Information Science and Technology,* 72**,** 885 - 900.

Li, W., C.-Y. Hsu & M. Hu (2021) Tobler's First Law in GeoAI: A Spatially Explicit Deep Learning Model for Terrain Feature Detection under Weak Supervision. *Annals of the American Association of Geographers,* 111**,** 1887-1905.

Liao, J., Q. Liao, W. Wang, S. Shen, Y. Sun, P. Xiao, Y. Cao & J. Chen (2023) Quantifying and mapping landscape value using online texts: A deep learning approach. *Applied Geography,* 154**,** 102950.

Lindell, M., P. Murray-Tuite, B. Wolshon & J. Baker. 2018a. *Large-Scale Evacuation: The Analysis, Modeling, and Management of Emergency Relocation from Hazardous Areas*.
---. 2018b. Who Leaves and Who Does Not: The Analysis, Modeling, and Management of Emergency Relocation from Hazardous Areas. 67-99.
Liu, X. H., B. Kar, C. Y. Zhang & D. M. Cochran (2019a) Assessing relevance of tweets for risk communication. *International Journal of Digital Earth,* 12**,** 781-801.
Liu, Y., M. Ott, N. Goyal, J. Du, M. Joshi, D. Chen, O. Levy, M. Lewis, L. Zettlemoyer & V. Stoyanov (2019b) RoBERTa: A Robustly Optimized BERT Pretraining Approach. *ArXiv,* abs/1907.11692.
Martin, M. E. & N. Schuurman (2020) Social Media Big Data Acquisition and Analysis for Qualitative GIScience: Challenges and Opportunities. *Annals of the American Association of Geographers,* 110**,** 1335-1352.
Martín, Y., S. L. Cutter & Z. Li (2020) Bridging Twitter and Survey Data for Evacuation Assessment of Hurricane Matthew and Hurricane Irma. *Natural Hazards Review,* 21**,** 04020003.
Miyazaki, H., M. Nagai & R. Shibasaki (2015) Reviews of Geospatial Information Technology and Collaborative Data Delivery for Disaster Risk Management. *Isprs International Journal of Geo-Information,* 4**,** 1936-1964.
Mohanty, S. D., B. Biggers, S. Sayedahmed, N. Pourebrahim, E. B. Goldstein, R. Bunch, G. Q. Chi, F. Sadri, T. P. McCoy & A. Cosby (2021) A multi-modal approach towards mining social media data during natural disasters-A case study of Hurricane Irma. *International Journal of Disaster Risk Reduction,* 54.
Naaz, S., Z. Ul Abedin & D. R. Rizvi (2021) Sequence Classification of Tweets with Transfer Learning via BERT in the Field of Disaster Management. *Eai Endorsed Transactions on Scalable Information Systems,* 8.
Neely, S. R. & M. Collins (2018) Social Media and Crisis Communications: A Survey of Local Governments in Florida. *Journal of Homeland Security and Emergency Management,* 15.
Ng, M., R. Diaz & J. Behr (2016) Inter- and intra-regional evacuation behavior during Hurricane Irene. *Travel Behaviour and Society,* 3**,** 21-28.
Regnier, E. D. (2008) Public Evacuation Decisions and Hurricane Track Uncertainty. *Manag. Sci.,* 54**,** 16-28.
Scheele, C., M. Z. Yu & Q. Y. Huang (2021) Geographic context-aware text mining: enhance social media message classification for situational awareness by integrating spatial and temporal features. *International Journal of Digital Earth*.
Senkbeil, J. C., D. M. Brommer, P. G. Dixon, M. E. Brown & K. Sherman-Morris (2010) The perceived landfall location of evacuees from Hurricane Gustav. *Natural Hazards,* 54**,** 141-158.
Shi, W. (2021) Implementing Social Media - Practical Reflections from County Governments During Hurricane Matthew. *International Journal of Public Administration in the Digital Age*.
Sit, M. A., C. Koylu & I. Demir (2019) Identifying disaster-related tweets and their semantic, spatial and temporal context using deep learning, natural language processing and spatial analysis: a case study of Hurricane Irma. *International Journal of Digital Earth,* 12**,** 1205-1229.
Spinsanti, L. & F. Ostermann (2013) Automated geographic context analysis for volunteered information. *Applied Geography,* 43**,** 36-44.

Tan, M. J. & C. Guan (2021) Are people happier in locations of high property value? Spatial temporal analytics of activity frequency, public sentiment and housing price using twitter data. *Applied Geography,* 132**,** 102474.

Vaswani, A., N. Shazeer, N. Parmar, J. Uszkoreit, L. Jones, A. N. Gomez, Ł. Kaiser & I. Polosukhin. 2017. Attention is all you need. In *Proceedings of the 31st International Conference on Neural Information Processing Systems*, 6000–6010. Long Beach, California, USA: Curran Associates Inc.

Veil, S., T. Buehner & M. Palenchar (2011) A Work-In-Process Literature Review: Incorporating Social Media in Risk and Crisis Communication. *J of contingencies and crisis management,* 19.

Vongkusolkit, J. & Q. Y. Huang (2021) Situational awareness extraction: a comprehensive review of social media data classification during natural hazards. *Annals of GIS,* 27**,** 5-28.

Wang, W. & K. Stewart (2015) Spatiotemporal and semantic information extraction from Web news reports about natural hazards. *Computers Environment and Urban Systems,* 50**,** 30-40.

Wang, Z. & X. Ye (2019) Space, time, and situational awareness in natural hazards: a case study of Hurricane Sandy with social media data. *Cartography and Geographic Information Science,* 46**,** 334-346.

Wolf, T., L. Debut, V. Sanh, J. Chaumond, C. Delangue, A. Moi, P. Cistac, T. Rault, R. Louf, M. Funtowicz & J. Brew. 2020. Transformers: State-of-the-Art Natural Language Processing. In *EMNLP*.

Wu, Y., M. Schuster, Z. Chen, Q. V. Le, M. Norouzi, W. Macherey, M. Krikun, Y. Cao, Q. Gao, K. Macherey, J. Klingner, A. Shah, M. Johnson, X. Liu, L. Kaiser, S. Gouws, Y. Kato, T. Kudo, H. Kazawa, K. Stevens, G. Kurian, N. Patil, W. Wang, C. Young, J. R. Smith, J. Riesa, A. Rudnick, O. Vinyals, G. S. Corrado, M. Hughes & J. Dean (2016) Google's Neural Machine Translation System: Bridging the Gap between Human and Machine Translation. *ArXiv,* abs/1609.08144.

Wukich, C. (2016) Government Social Media Messages across Disaster Phases. *Journal of Contingencies and Crisis Management,* 24**,** 230-243.

Yu, M. Z., Q. Y. Huang, H. Qin, C. Scheele & C. W. Yang (2019) Deep learning for real-time social media text classification for situation awareness - using Hurricanes Sandy, Harvey, and Irma as case studies. *International Journal of Digital Earth,* 12**,** 1230-1247.

Zhao, B. (2017) Web scraping. *Encyclopedia of big data***,** 1-3.

Zhou, B., L. Zou, A. Mostafavi, B. Lin, M. Yang, N. Gharaibeh, H. Cai, J. Abedin & D. Mandal (2022) VictimFinder: Harvesting rescue requests in disaster response from social media with BERT. *Computers, Environment and Urban Systems,* 95**,** 101824.

Zhou, X., M. Wang & D. Li (2017) From stay to play – A travel planning tool based on crowdsourcing user-generated contents. *Applied Geography,* 78**,** 1-11.

Appendix A. Sample mandatory evacuation notice text data and their BERT classification probability and predictive result.

| Case Number | Geog. Area | Text Data | BERT Prob. | Classified Correctly |
|---|---|---|---|---|
| 75 | Lafourche, Louisiana | Parish President Archie Chaisson has issued a Mandatory Evacuation for residents and businesses south of the Leon Theriot Floodgate in Golden Meadow and other low-lying areas of Lafourche Parish, effective 6pm today. | 0.9948 | Yes |
| 528 | St. Mary, Louisiana | St. Mary Parish Sheriff's Office 1 hr *****HURRICANE LAURA UPDATE AND INFORMATION***** . Sheriff Blaise Smith and the deputies of the SMPSO were just updated by the National Weather Service in Lake Charles. . The information regarding St. Mary … More Parish for Hurricane Laura is as follows: . We are under a Tropical Storm Warning as strong tropical storm winds with gusts up to 58 mph are expected in the parish. Winds are expected to start Wednesday afternoon and continue into Thursday afternoon (16-20 hours). There will be intermittent hurricane-force wind gusts Thursday morning. . We are also under a storm surge warning. . ****MANDATORY EVACUATION for those in these areas**** - South of the Intracoastal Waterway - South of Highway 83 & Cypremort Point Road THIS IS NOT AN EVACUATION FOR THE WHOLE PARISH. It is just for those low lying areas that will be affected by the storm surge. . ****CURFEW FOR ST MARY PARISH**** . A curfew for St. Mary Parish has been established from 9 pm to 6 am. This curfew will stay in effect until further notice. . #Laura | 0.8964 | Yes |
| 175 | Galveston, Texas | Galveston County Judge Mark Henry has upgraded the voluntary evacuation for the Bolivar Peninsula to a mandatory evacuation. A voluntary evacuation has also been issued for the Bay Shore areas of San Leon and … More Bacliff and the unincorporated area of Freddiesville. All Galveston County residents should be taking preparations in advance of this storm. This includes preparing your homes, ensuring your disaster kit is stocked, and heeding all warnings and evacuation orders. For more information, visit www.GCOEM.org. | 0.7870 | Yes |
| 1214 | New Haven, Connecticut | Town of Madison, CT - Government Message from First Selectman Fillmore McPherson with a Hurricane Sandy update as of Sunday afternoon: | 0.6511 | Yes |

|  |  | Heavy winds from the storm are expected to start around midnight tonight and to last through Tuesday. This is a much longer length of time than we have seen in previous storms, and we will experience at least three high tides of near catastrophic proportions. Flooding along the coast and some inland areas will be much worse than we saw with Irene last year. At noon today I issued a mandatory evacuation notice for the low lying areas. I requested people to leave by sundown today. By far the best choice of where to go is to friends or relatives north of I-95 or out of town. |  |  |
|---|---|---|---|---|
| **725** | Montgomery, Texas | Record levels of water are being released from Lake Conroe Dam. Flooding imminent in some areas. The City of Conroe will be evacuating McDade Estates. Other neighborhoods will be evacuated by the County. | 0.4126 | No |
| **1283** | Indian River, Florida | Throughout the remainder of the day Wednesday, parts of Indian River County including the barrier island, are under a voluntary evacuation. As of 8 a.m. Thursday, a mandatory evacuation order will be issued for the barrier island, mobile/manufactured homes, low-lying areas and substandard housing, according to the Indian River County Sheriff's Office. | 0.1673 | No |